\documentclass[twocolumn,showpacs,preprintnumbers,amsmath,amssymb]{revtex4}
\usepackage{graphicx,epsfig}
\usepackage{dcolumn}
\usepackage{bm}

\def\rop{r_{\hbox{\tiny{opt}}}}
\def\vop{v_{\hbox{\tiny{opt}}}}
\def\bea{\begin{eqnarray}}
\def\eea{\end{eqnarray}}
\def\bi{\begin{itemize}}
\def\ei{\end{itemize}}

\def\2te{2{\theta}}

\def\'#1{\ifx#1i\accent19\i\else\accent19#1\fi}
\begin{document}
%
%
\title{Hydrodynamical description of Galactic Dark Matter
\footnote{Contribution to Proceeedings of the {\it IV Taller 
de la Divisi\'on de Gravitaci\'on y F{\'\i}sica Matem\'atica 
de la Sociedad Mexicana de F{\'\i}sica ({\bf DGyFM-SMF})}. Chapala, 
Jalisco, M\'exico November 25 - November 30, 2001. This paper is 
based on the Ref. \cite{nosotros}.} }
\author{Luis G. Cabral-Rosetti}
 \email{luis@nuclecu.unam.mx}
\author{Dar{\'\i}o N\'u\~nez}
 \altaffiliation[Also at ]{Center for Gravitational Physics and 
                           Geometry, Penn State University, University 
                           Park, PA, 16802, USA.}
 \email{nunez@nuclecu.unam.mx, nunez@gravity.phys.psu.edu}
 \author{Roberto A. Sussman}
 \email{sussman@nuclecu.unam.mx}
\affiliation{Instituto de Ciencias Nucleares, \\
             Universidad Nacional 
             Aut\'onoma de M\'exico (ICN-UNAM).\\
             Apartado Postal 70-543, 94510 M\'exico, D.F., M\'exico.}
\author{Tonatiuh Matos}
 \email{tmatos@fis.cinvestav.mx}
\affiliation{Departamento de F{\'\i}sica, Centro de Investigaci\'on y 
             Estudios Avanzados del IPN,\\
             Apartado Postal 14-740, M\'exico D. F., M\'exico.}
%
%
\begin{abstract}
We consider simple hydrodynamical models of galactic dark matter in which 
the galactic halo is a self-gravitating and self-interacting gas that 
dominates the dynamics of the galaxy. Modeling this halo as a sphericaly 
symmetric and static perfect fluid satisfying the field equations of General 
Relativity, visible barionic matter can be treated as  ``test particles'' 
in the geometry of this field.  We show that the assumption of an empirical 
``universal rotation curve'' that fits a wide variety of galaxies is 
compatible, under suitable approximations, with state variables 
characteristic of a non-relativistic Maxwell-Boltzmann gas that becomes 
an  isothermal sphere in the Newtonian limit. Consistency criteria lead
to a minimal bound for particle masses in the range 
$30 \,\hbox{eV} \alt m \alt 60 \,\hbox{eV}$ and to a constraint between 
the central 
temperature and the particles mass. The allowed mass range includes popular 
supersymmetric particle candidates, such as the neutralino, axino and 
gravitino, as well as lighter particles ($m\sim$ keV) proposed by
numerical N-body simulations associated with self-interactive ``cold''
and ``warm'' dark matter structure  formation theories.
\end{abstract}
\maketitle

\section{Introduction}

The presence of large amounts of dark matter at the galactic lengthscale
is already an established fact. It is currently thought that this dark
matter is made of relique self-gravitating gases which are labeled as 
``cold''(CDM) or ``hot'' (HDM), depending on the relativistic or 
non-relativistic nature of the particles energetic spectrum at their 
decoupling from the cosmic mixture \cite{Trimble}, \cite{tbook_1}, 
\cite{tbook_2}.  HDM scenarios are not favoured, as they seem to be
incompatible  with current theories of structure formation \cite{tbook_1},
\cite{tbook_2}, \cite{dm_scenarios}. CDM, usualy
examined within a newtonian framework, can be considered as non-interactive
(collisionless particles) or self-interactive \cite{old_cdm_theo}. N-body
numerical simmulations are often used for modeling CDM gases
\cite{old_cdm_nbody}, \cite{nbody_1}, \cite{nbody_2}, \cite{nbody_3}.
However, in recent numerical simulations (see \cite{nbody_1}, \cite{nbody_2},
\cite{nbody_3}) non-interactive CDM models present the following
discrepancies with observations at the galactic scale \cite{cdm_problems_1},
\cite{cdm_problems_2}: (a) the ``substructure problem'' related to excess 
clustering on sub-galactic scales, (b) the ``cusp problem'' characterized by 
a monotonic  increase of density towards the center of halos, leading to 
excessively  concentrated cores. In order to deal with these problems, the 
possibility of self-interactive dark matter has been considered, so that 
nonzero  pressure or thermal effects can emerge, thus leading to 
self-interactive models of CDM ({\it i.e.} SCDM) \cite{scdm_1}, 
\cite{scdm_2}, \cite{scdm_3}, \cite{scdm_4}, \cite{scdm_5} and ``warm'' 
dark matter (WDM) models \cite{wdm_1}-\cite{wdm_6} that challenges the 
duality CDM vs. HDM. Other proposed dark matter sources consist replacing 
the gas of particles approach by scalar fields \cite{sfe_1}, \cite{sfe_2} 
and even more ``exotic'' sources\cite{sfe_3}.

Whether based on SCDM or WDM, current theories of structure formation
point towards dark matter characterized by particles having a mass of the
order of at least keV's (see \cite{scdm_1}-\cite{wdm_6}), thus suggesting that
massive  but light particles, such as electron neutrinos and axions (see
Table~1), should be eliminated as primary dark matter candidates (though 
there is no reason to assume that these particles would be absent in galactic 
halos). Of all possible weakly interactive massive particles (WIMPS), 
complying with the required mass value of relique gases, only the massive 
Neutrinos (the muon or tau neutrinos), have been detected, whereas other 
WIMPS  (neutralino, gravitino, photino, sterile neutrino, axino, etc.) are
speculative.  See \cite{Pal}, \cite{SNO} \cite{Ellis} and Table~1 for a
list of candidate  particles and appropriate references.

In this paper we develop an alternative description of galactic DM. Since
the dark matter halo constitutes about 90 \% of the galactic mass, we 
consider the galactic garvitational field as a spacetime whose sole,
self-gravitating, source is this halo, described as a perfect fluid.
Assuming this galactic spacetime to be static and sphericaly symmetric, the 
barionic dark matter can be though of as test particles following stable and 
circular geodesics of this spacetime curvature. Since the tangential velocity,
$v$, along these geodesics can be calculated for such a spacetime, we can 
express the field equations in terms of of $v$. However, the velocity profile 
$v=v(r)$ has been observed for a wide range of galaxies, leading to 
``Universal Rotatition Curves'' (URC's) that provide an empiric fit to these 
rotation velocities\cite{urc_2}. Therefore, by inserting the empiric formula 
for the URC derived by Persic and Salucci \cite{urc_1} into the field 
equations (given in terms of $v(r)$), we obtain a constraint on one of the 
metric coefficients. Once this constraint is solved, we can obtain the state
variables characterizing the galactic dark fluid associated with this URC. We
solve this constraint assuming that the velocities are non-relativistic, thus
expanding around $v_0/c\ll 1$, where $v_0$ is the terminal velocity associated
with the flattening of the URC.

\section{Field Equations}

Considering the line element of a static spherically symmetric space time

\begin{equation}
\begin{array}{c}
\displaystyle
ds^2 = -A^2(r)\,c^2\,dt^2 + \frac{d\,r^2}{1-2\,M(r)/r}
\\[0.5cm]
\displaystyle
+ r^2\,\left(d\theta^2+\sin^2\theta\,d\varphi^2\right)\ , 
\label{eq:ele0}
\end{array}
\end{equation}

\noindent
the tangential velocity of test particles along stable circular geodesic
orbits can be expressed in terms of the metric coefficients as

\begin{equation} 
\frac{V^2}{c^2} \ \equiv \ v^2(r) \ = \
{\frac{{r\,A'}}{{A}}}\ .
\label{eq:vt} 
\end{equation}

\noindent
Becomes a dynamical variable replacing $A(r)$. Assuming as source of
(\ref{eq:ele0}) a perfect fluid momentum energy tensor: 
$T^{ab}=(\rho+p)\,u^au^b+p\,g^{ab}$, with
$u^a=A^{-1}\,\delta^a\,_{ct}$, the following field equations in terms of
(\ref{eq:vt}) become

\begin{equation}
\begin{array}{c}
\displaystyle
M' + {\frac {\left (-3-5\,v^2+4\,vv'\,r+2\,{v}^{4}\right )M}
{r\left (1+{v}^{2}\right )}}
\\[0.5cm]
\displaystyle
- {\frac {v\left (-2\,v+2\,v'\,r+{v}^{3} \right )}{1+{v}^{2}}} = 0\ .
\label{eq:const}
\end{array}
\end{equation}

\begin{equation}
\kappa\,p\ = \ 2\,{\frac {-M-2\,M{v}^{2}
+{v}^{2}r}{{r}^{3}}}\ , 
\label{eq:p}
\end{equation}

\begin{equation}
\begin{array}{c}
\displaystyle
\kappa\,\rho =  {\frac {\left [-8\,vv'\,r-2\,\left (2\,{v}^{2}+1
\right )\left ({v}^{2}-3\right )\right ]M}{{r}^{3}\left (1+{v}^{2}\right )}}
\\[0.5cm]
\displaystyle
+\ {\frac {4\,{r}^{2}vv'+2\,{v}^{2 }\left
(-2+{v}^{2}\right)r}{{r}^{3}\left (1+{v}^{2}\right )}}\ ,
\label{eq:rho}
\end{array}
\end{equation}

\noindent
where $\kappa =8\pi G/c^4$ and a prime denotes derivative with respect to
$r$. Writing the field  equations in terms of the orbital velocity, $v$, 
provides a useful insight into how an (in principle) observable quantity 
relates to spacetime curvature  and with physical quantities (state variables)
which characterize the source  of spacetime.

\section{Thermodynamics}

If we assume that the self gravitating ideal ``dark'' gas  exists in physical
conditions far from those in which the quantum properties of the gas
particles are relevant, we would be demanding that these particles comply with
Maxwell-Boltzmann (MB) statistics. Following  \cite{Landau}, the condition
that justifies an MB distribution is given by

\begin{equation}
\frac{n\, \hbar^3}{(m\, k_{_B}\, T)^{3/2}} \ll 1 \ ,
\label{eq:landau}
\end{equation}

\noindent
where $n$, $T$, $\hbar$ and $k_{_B}$ are, respectively, the particle number
density, absolute temperature, Planck's and Boltzmann's constants. If the
constraint (\ref{eq:landau}) holds and we further assume thermodynamical
equilibrium and non-relativistic conditions, the ideal dark gas must satisfy
the equation of state of a non-relativistic monatomic ideal gas

\begin{equation}
\rho \ = \ mc^2\,n \ + \ \frac{3}{2}\,n\,k_{_B}T\ \ \ , \ \ \  
p  = n\,k_{_B}T,
\label{eq:mb}
\end{equation}

\noindent
whose macroscopic state variables  can be obtained from a MB distribution
function under an equilibrium Kinetic theory approach (the non-relativistic
and non-degenerate limit of the J\"uttner distribution) \cite{RKT}.  An
equilibrium MB distribution restricts the geometry of spacetime \cite{rund},
resulting in the existence of a timelike Killing vector field
$\beta^a=\beta\,u^a$, where $\beta\equiv mc^2/k_{_B}T$, as well as the 
following relation (Tolman's law) between the 4-acceleration and the 
temperature gradient

\begin{equation}
\dot u_a \ + \ h_a^b\,(\ln T)_{,b} \ = \ 0\ \ \ \ ,\ \ \ \ \
h_a^b = u_a u^b + \delta_a^b\ ,
\label{eq:tolma}
\end{equation}

\noindent
leading to

\begin{equation}
\frac{A'}{A} \ + \ \frac{T'}{T} \ = \ 0 \qquad
\Rightarrow \qquad T \ \propto \ A^{-1}\ .
\label{eq:Ttolman}
\end{equation}

The particle number density $n$ trivialy satisfies the conservation law
$J^a\,_{;a}=0$ where $J^a=n\,u^a$, thus the number of dark particles is
conserved. Notice that given (\ref{eq:p}) and (\ref{eq:rho}), the equation
of state (\ref{eq:mb}) and the temperature from the Tolman law 
(\ref{eq:Ttolman}), we have two different expressions for $n$

\begin{equation}
n \ = \ \frac{p}{k_{_B}T} \ \propto \ p\,A,
\label{eq:n1}
\end{equation}

\begin{equation}
\begin{array}{c}
\displaystyle
n \ = \ \frac{1}{mc^2}\,\left[\rho - \frac{3}{2}\,p\right]
\\[0.5cm]
\displaystyle
= {\frac {\left [-8\,vv'\,r+\left ({v}^{2} + 9\right)\left 
(2\,{v}^{2} + 1\right )\right ]M}{\kappa\,mc^2\,{r}^{3}
\left (1+{v}^{2}\right )}}
\\[0.5cm]
\displaystyle
+\ {\frac {+4\,vv'\,{r}^{2}-{v}^{2}\left (7+{v}^{2}\right)r}
{\kappa\,mc^2\,{r}^{3}\left (1+{v}^{2}\right )}}\ .
\end{array}
\label{eq:n2}
\end{equation}

The quantity $mc^2\,n$ in (\ref{eq:n2}) follows directly from equations
(\ref{eq:p}) and (\ref{eq:rho}), while $n$ in (\ref{eq:n1}) also follows
from $p$ in (\ref{eq:p}) with $A\propto \exp[\int{(v^2/r)dr}]$. Consistency 
requires that (\ref{eq:n1}) and (\ref{eq:n2}) yield the same expression for 
$n$.

\section{Dark fluid hydrodynamics}

We shall assume for $v^2$ the empiric dark halo rotation velocity law given
by Persic and Salucci \cite{urc_1}, \cite{urc_2}

\begin{equation}
v^2 \ = \ \frac{v_0^2\,x^2}{a^2+x^2}\ \ \ ,\ \ \  x \ \equiv \ \frac{r}{\rop}
\label{eq:v}
\end{equation}

\noindent
where $\rop$ is the ``optical radius'' containing 83 \% of the galactic 
luminosity, whereas the empiric parameters $a$ and $v_0$, respectively, the
ratio of ``halo core radius'' to $\rop$ and the ``terminal'' rotation velocity,
depend on the galactic luminosity. For spiral galaxies we have:
$v_0^2=\vop^2(1-\beta)(1+a^2)$, where $\vop=v(\rop)$  and the best fit to
rotation curves is obtained for: $a=1.5\,(L/L_*)^{1/5}$ and
$\beta=0.72+0.44\log_{10}(L/L_*)$, where $L_*=10^{10.4} L_\odot$. The range 
of these parameters for spiral galaxies is 
$125 \,\hbox{km/sec} \leq v_0 \leq 250 \,\hbox{km/sec}$ and 
$0.6 \leq a \leq 2.3 $.

\noindent
Inserting (\ref{eq:v}) into (\ref{eq:vt}) and (\ref{eq:const}) we obtain

\begin{equation}
A \ = \ \left[1+x^2\right]^{v_0^2/2} \
\Rightarrow \ T \ = \ T_c
\,\left[1+x^2\right]^{-v_0^2/2}\ ,
\label{eq:AT}
\end{equation}

\begin{equation}
\begin{array}{c}
\displaystyle
M \ = \ \frac{(v_0^2-2)(a^2+x^2)^{2-v_0^2}\,v_0^2\,x^3}
{\left[a^2+(1+v_0^2)x^2\right]^{2/(1+v_0^2)}}\,
\\[0.5cm]
\displaystyle
\times\ \rop\,
\int{\frac{\left[a^2+(1+v_0^2)x^2\right]^{(1-v_0^2)/(1+v_0^2)}\,x\,dx}
{(a^2+x^2)^{3-v_0^2}}}\ ,
\end{array}
\label{eq:M2}
\end{equation}

\noindent
where $T_c=T(0)$ and we have set an integration constant to zero in order
to comply with the consistency requirement that $v_0=0$ implies flat spacetime
($A=1,\,M=0$). Since the velocities of rotation curves are newtonian,
$v_0\ll c$ (typical values are $v_0/c\approx 0.5\times 10^{-3}$), instead of
evaluating (\ref{eq:M2}) we will expand this  quadrature around $v_0/c$ (in
order to keep the notation simple, we write $v_0$ instead of $v_0/c$). This 
yields

\begin{figure}[htb]
\centerline{
\epsfig{figure=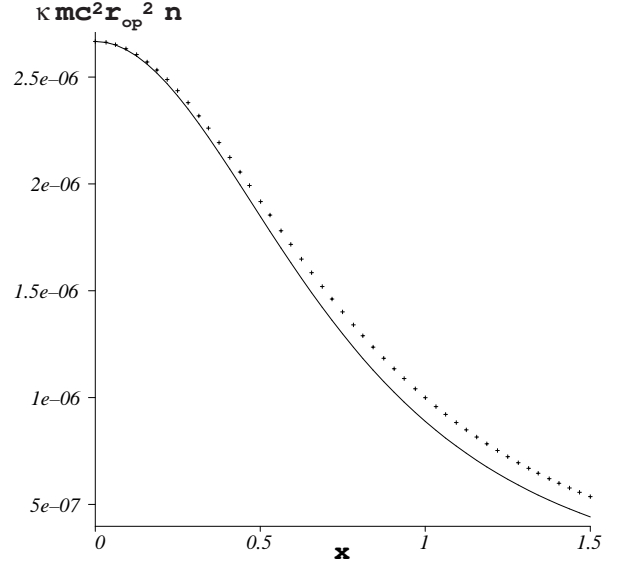,width=8cm}}
\caption{{\it Comparison of $n$ obtained from (\ref{eq:n11}) and 
(\ref{eq:n22}). This plot displays the adimensional quantity 
$\kappa m c^2 r_{op}^2\,n$, as a function of $x$, obtained from truncating 
the right hand sides of (\ref{eq:n11}) (solid curve) and (\ref{eq:n22}) 
(dotted curve with crosses) up to fourth order in $v_0/c$ and  assuming 
the consistency condition (\ref{eq:consist}). Since we are using an empiric 
law for observed galactic rotation curves (the URC given by (\ref{eq:v})), 
the fact that these two expressions for $n$ are so close to each other 
provides an empirical justification for the compatibility between MB 
distribution and these rotation curves}.}
\label{fig1}
\end{figure}

\begin{equation}
\begin{array}{c}
\displaystyle
M \ = \ \frac{x^3\,\rop}{a^2+x^2}\,v_0^2\,
\Bigg[1-\frac{5x^2+2a^2}{2(a^2+x^2)}\,v_0^2
\\[0.5cm]
\displaystyle
+\ \frac{12x^4+11a^2x^2+3a^4}{2(a^2+x^2)^2}\,v_
0^4 + {\cal O}(v_0^6)\Bigg]\ ,
\end{array}
\label{eq:M3}
\end{equation}

\noindent
we obtain the expanded forms of $\rho$ and $p$ by inserting (\ref{eq:v}) and 
(\ref{eq:M2}) into (\ref{eq:p}) and (\ref{eq:rho}) and then expanding around 
$v_0$, leading to

\begin{equation}
\begin{array}{c}
\displaystyle
\kappa\,\rho\,\rop^2 \ = \
\frac{2(3a^2+x^2)}{(a^2+x^2)^2}\,v_0^2
\\[0.5cm]
\displaystyle
-\ \frac{5x^4+23a^2x^2+6a^4}{(a^2+x^2)^3}\,v_0^4
+ {\cal O}(v_0^6)\ ,
\label{eq:rho3}
\end{array}
\end{equation}

\begin{equation}
\begin{array}{c}
\displaystyle
\kappa\,p\,\rop^2 \ = \ \frac{2a^2+x^2}{(a^2+x^2)^2}\,v_0^4 
\\[0.5cm]
\displaystyle
-\ \frac{2x^4+7a^2x^2+3a^4}{(a^2+x^2)^3}\,v_0^6 + {\cal O}(v_0^8)\ ,
\label{eq:p3}
\end{array}
\end{equation}

\noindent
while the expanded form for $T$ follows from (\ref{eq:AT})

\begin{figure}[htb]
\centerline{
\epsfig{figure=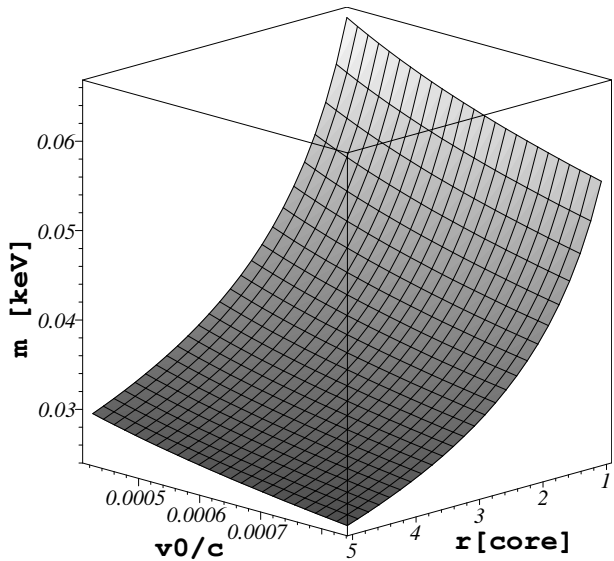,width=8cm}}
\caption{{\it Minimal mass for which the Maxwell-Boltzmann
distribution is applicable. This graph displays $m$ (in keV's) as a 
function of $v_0/c$ and $b=a\,x$, respectively, the terminal velocity and 
`halo core radius' associated with the URC given in (\ref{eq:v}). Assuming 
typical ranges for spiral galaxies: 
$125\,\hbox{km/sec} \leq v_0 \leq 250\,\hbox{km/sec}$ and 
$1\,\hbox{kpc} \leq b \leq 5\,\hbox{kpc}$, we obtain masses in the range of 
$30 \,\hbox{eV} \leq m \leq 60 \,\hbox{eV}$ that follow from the right hand 
side of the relation (\ref{aplicab}), providing the criterion for 
applicability of the Maxwell-Bolzmann distribution. Dark matter particle 
candidates complying with an MB distribution must have much larger mass 
than the plotted values\,  $30 \,\hbox{eV} \leq m \leq 60 \,\hbox{eV}$.}}
\label{fig2}
\end{figure}

\begin{equation}
\begin{array}{c}
\displaystyle
T \ = \ T_c\,\Bigg[1-\frac{1}{2}\,\ln\left(1+x^2\right),v_0^2 
\\[0.5cm]
\displaystyle
+\ \frac{1}{8}\,\ln^2\left(1+x^2\right)\,v_0^4
- {\cal O}(v_0^6)\Bigg]\ .
\end{array}
\end{equation}

In order to compare $n$ obtained from (\ref{eq:n1}) and (\ref{eq:n2}), we
substitute (\ref{eq:v}) and (\ref{eq:M2}) into (\ref{eq:n2}) and expand
around $v_0$, leading to

\begin{equation}
\begin{array}{c}
\displaystyle
n\,\rop^2 \ = {{1}\over{\kappa\,m\,c^2}}\,
\Bigg[\frac{2(3a^2+x^2)}{(a^2+x^2)^2}\,v_0^2
\\[0.5cm]
\displaystyle
-\ \frac{18x^4+55a^2x^2+13a^4}{2(a^2+x^2)^3}\,v_0^4 
+ {\cal O}(v_0^6)\Bigg],
\label{eq:n11}
\end{array}
\end{equation}

\noindent
while $n$ in (\ref{eq:n1}) follows by substituting (\ref{eq:M2}) into
(\ref{eq:p3}), using $T$ from (\ref{eq:AT}) and then expanding around 
$v_0$. This yields

\begin{equation}
\begin{array}{c}
\displaystyle
n\,\rop^2 \ = {{1}\over{\kappa\,k_{_B}T_c}}\Bigg[
\frac{2a^2+x^2}{(a^2+x^2)^2}\,v_0^4
\\[0.5cm]
\displaystyle
+\ \frac{(2a^2+x^2)(a^2+x^2)
\ln\left(1+x^2\right)}{2(a^2+x^2)^3}\,v_0^6
\\[0.5cm]
\displaystyle
-\ \frac{2(a^2+2x^2)(3a^2+x^2)}{2(a^2+x^2)^3}\,v_0^6
+ {\cal O}(v_0^8) \Bigg]\ . 
\label{eq:n22}
\end{array}
\end{equation}

Since $v_0/c\ll 1$, a reasonable approximation is obtained if the leading 
terms of $n$ from (\ref{eq:n11}) and (\ref{eq:n22}) coincide. By looking at 
these equations, it is evident that this consistency requirement implies

\begin{equation}
\frac{1}{2}\,mv_0^2 \ = \ \frac{3}{2}\,k_{_B}T_c,
\label{eq:consist}
\end{equation}

\noindent
where $v_0$ denotes a velocity ($\hbox{cm}/\hbox{sec} $) and not the
adimensional ratio $v_0/c$. Since higher order terms in $v_0/c$ have a minor
contribution, the two forms  of $n$ are approximately equal. This is shown in
Figure~1 displaying the adimensional quantity $\kappa\, m c^2 n\,\rop^2$  from
(\ref{eq:n11}) and (\ref{eq:n22}) as functions of $x$ for typical values
$v_0/c=0.0006,\,a=1$ and eliminating $T_c$ with (\ref{eq:consist}).
Equation (\ref{eq:M2}) shows how ``flattened'' rotation curves, as obtained
from the empiric form (\ref{eq:v}), lead to $M\propto r^3$ for $r\approx 0$  
and $M\propto r$ for large $r$. Equations (\ref{eq:M3}) to  
(\ref{eq:consist}) represent a relativistic generalization of the 
``isothermal sphere'' that follows as the newtonian limit of an ideal 
Maxwell-Boltzmann gas characterized by $\rho\approx mc^2n$,\,$p\ll \rho$ and 
$T\approx T_c$. In fact, using newtonian hydrodynamics we would have obtained
only the leading terms of equations (\ref{eq:M3}) to (\ref{eq:consist}). It
is still interesting to find out that the isothermal sphere can be obtained
from General Relativity in the limit $v_0/c\ll 1$ by demanding that rotation
curves have a form like (\ref{eq:v}). The total mass of the galactic halo,
usualy given as $M$ evaluated at the radius $r=r_{200}$ (the radius at which 
$\rho$ is 200 times the mean cosmic density). Assuming this density to be 
$\approx 10^{-29}\,\hbox{gm}/\hbox{cm}^3$ together with typical values 
$v_0=200 \hbox{km}/\hbox{sec}$ and $a=1$  yields $r_{200}\approx 150$ kpc.
Evaluating $M$ at this values yields about $10^{17}\,M_\odot$, while $M$ 
evaluated at a typical ``optical radius'' $r=15$ kpc leads to about 
$10^{12}\,M_\odot$, an order of magnitude larger than the  galactic mass due 
to visible matter.

\section{Discussion.}

So far we have found a reasonable approximation for galactic dark matter
to be described by a self gravitating Maxwell-Boltzmann gas, under the
assumption of the empiric rotation velocity law (\ref{eq:v}). The following
consistency relations emerge from equations (\ref{eq:n11}), (\ref{eq:n22})
and (\ref{eq:consist})

\begin{equation} n_c \ \approx \ \frac{3\,v_0^2}{4\pi G\,m\,a^2\,\rop^2},\qquad T_c \
\approx
\ \frac{m\,v_0^2}{3\,k_{_B}}
\label{eq:nTc}\end{equation}

\noindent
hence, bearing in mind that $n\leq n_c$ and $T\approx T_c$, the condition
(\ref{eq:landau}) for the validity of the MB distribution together with
(\ref{eq:nTc}) yields the condition

\begin{equation} m \ \gg \ \left[\frac{3^{5/2}\,\hbar^3}{4\pi
G\,a^2\,\rop^2\,v_0}\right]^{1/4},\label{aplicab}
\end{equation}

\noindent
a criteria of aplicability of the MB distribution that is entirely given in
terms of $m$, the fundamental constants $G,\,\hbar $ and the empiric
parameters $v_0$ and $a\,\rop$ (the ``terminal'' rotation velocity and the
``core radius'') \cite{nosotros}. For dark matter dominated galaxies (spiral 
and low surface brightness (LSB)) \cite{urc_1} these parameters have a small
variation  range: $\rop \approx 15\,\hbox{kpc} $, $0.6 \,  \leq a \leq \, 2.3$
and $125\, \hbox{km/sec}  \leq v_0 \leq 300\, \hbox{km/sec}$, the constraint 
(\ref{aplicab}) does provide a tight estimate of the minimal value for the 
mass of the particles under  the assumption that these particles form a self 
gravitating ideal dark gas  complying with MB statistics. As shown in 
Figure~2, this minimal value lies  between $30$ and $60$ $eV$, thus implying 
that appropriate particle candidates must have a much larger mass than this 
figure \cite{nosotros}. This minimal bound excludes, for instance, light mass 
thermal particles such as the electron neutrino ($m_{\nu_e} < 2.2$ eV). The axion 
is also very light ($m_A \approx 10^{-5}$ $eV$) but it is not a thermal relique and so we cannot study
it under the present framework. The currently accepted estimations of  cosmological bounds on the sum of
masses for the three active neutrino  species is about $24$ $eV$, a value that would apparently rule
out  all neutrino flavours. However, recent estimations of these cosmological  bounds have raised this
sum to about $1$
$keV$
\cite{sn2}, hence more massive neutrinos could also be accomodated as dark matter particle candidates. 
Estimates of masses of various particle candidates are displayed in Table~1.

\begin{table}
\caption{\label{tab:table1}
{\em Particle candidates for a MB Dark Matter gas}.}
\begin{ruledtabular}
\begin{tabular}{lcr}
\ \ SCDM/WDM&mass in keV&References\\
Light Candidates& & \\
\hline
Light Gravitino\hfill & $\sim 0.5$  & \cite{g1} \\
\ \ \ \ \ \ \ \ \ ''  & $\sim 0.75-1.5$ & \cite{g2,g3}\\
Sterile Neutrino & $\sim 2.6 - 5$ & \cite{ssn1}\\
\ \ \ \ \ \ \ \ \ '' & $< 40$ & \cite{ssn2}\\
\ \ \ \ \ \ \ \ \ '' & $1 - 100$ & \cite{ssn3}\\
Standard Neutrinos & $\sim 1$ & \cite{sn1,sn2}\\
Light Dilaton & $\sim 0.5$ & \cite{d1}\\
Light Axino & $\sim 100$ & \cite{a1}\\
Majoron & $\sim 1$ & \cite{m1,m2,m3}\\
Mirror Neutrinos & $\sim 1$ & \cite{mn1,mn2}\\
\end{tabular}
\begin{tabular}{lcr}
\ \ \ \ \ \ \ \ {CDM} & {mass in GeV} & {References} \\
{Heavy Candidates } & {} & {} \\
\hline
{Neutralino\hfill} & {$> 32.3$} & {\ \cite{abreu} } \\
{\hfill} & {$> 46$ } & {\cite{ellis} } \\
{Axino\hfill} & {$\sim 10$} & {\cite{covi,leszek}} \\ 
{Gravitino\hfill} & {${\buildrel <\over \sim }\ 100$} 
& {\cite{kawasaki} } \\ \hline
\end{tabular}
\end{ruledtabular}
\end{table}

\begin{figure}[htb]
\centerline{
\epsfig{figure=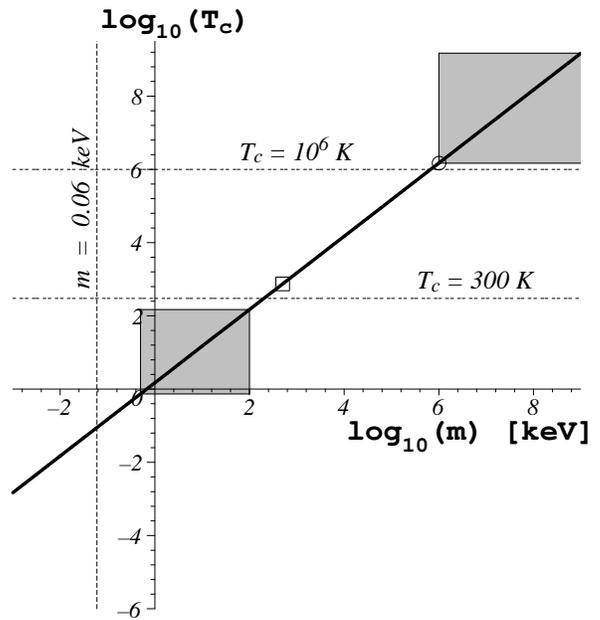,width=8cm}}
\caption{{\it Relation between particle mass and central
temperature. This graph displays the relation between $\log_{10}(T_c)$
(in K) and $\log_{10}(m)$ (in keV's) that follows from equation 
(\ref{eq:consist2}) for a terminal velocity $v_0=200\,\hbox{km/sec} $. 
Almost identical plots are obtained for other velocities in the observed 
range $125 \,\hbox{km/sec} \leq v_0 \leq 250 \,\hbox{km/sec}$. The circle 
and box symbols respectively denote the proton and electron mass yielding 
central temperatures of the order $T_c\approx 10^6,\,10^3 $ K. The central 
temperature for light particles in the range 
$0.5 \,\hbox{keV} \leq m \leq 100 \,\hbox{keV}$ is
less than $300$ K (rectangle in the left), while for massive sypersymmetric
particles in the range $1 \,\hbox{GeV} \leq m \leq 100 \,\hbox{GeV}$, we 
have $T_c$ as large as $10^9$ K (rectangle on the right). However, such high 
temperatures cannot rule out these weakly interactive particles as components 
of the dark matter MB gas}.}
\label{fig3}
\end{figure}

Since $T\approx T_c$, the consistency condition (\ref{eq:consist}) provides
the  following constraint on the temperature and particles mass of the dark
gas

\begin{equation}
\frac{m}{T_c} \ = \ \frac{3\,k_{_B}}{v_0^2} \ \approx 0.4\times 10^{3}\,
\frac{\hbox{eV}}{\hbox{K}},
\label{eq:consist2}
\end{equation}

\noindent
where we have taken $v_0 = 300\,\hbox{km/sec}$. Considering in
(\ref{eq:consist2}) the minimal mass range that follows from (\ref{aplicab}),
we would obtain gas temperatures consistent with the assumed typical
temperatures of relic gases: $T_c\approx 2 \,\,\hbox{to}\,\, 4\, \hbox{K}$.
However, since we have no way of infering a value for the temperature of the
ideal dark gas, we have no clear cut criterion for the estimate of a maximal
bound for this mass. If we assume that the ideal dark gas is made of electrons or barions, so that
$m=m_p$ or $m=m_e$, then condition (\ref{aplicab}) for applicability of the MB distribution is
certainly satisfied and  (\ref{eq:consist2}) implies a temperature of the
order of $T_c\approx 10^3$ K for electrons and $T_c\approx 10^6$ K for
barions. Obviuosly, barions or electrons at such a high temperatures would radiate and
certainly not remain unobservably ``dark''. However, as long as the interaction is weak and the
particles are not charged, we cannot rule out any other particle candidate only on the basis of the gas
temperature, even if this temperature is very high (see Figure~3) as in the case of massive
supersymmetric particles. As shown in Table~1, a wide range of weakly interactive particles can be
considered as possible main components of a MB dark gas, including 
popular supersymmetric particles (the neutralino), as well as hypothetical light particles
predicted by current literature based on WDM models of structure formation
\cite{tbook_1}, \cite{tbook_2}. The main  novelty of the present paper is the fact that it is based on
a  general  relativistic hydrodynamics, as opposed to numerical simulations 
\cite{nbody_1}-\cite{nbody_3}, newtonian or Kinetic Theory perturbative 
approaches (see \cite{scdm_1}-\cite{wdm_6}).

Finaly, the fact that we have obtained a minimal mass on the range $30-60$
$eV$, that seems to discriminate against very light thermal particles like the electron neutrino,  coincides
with the fact that these HDM particle candidates tend to be ruled out because of their inability to produce
sufficient matter clustering \cite{tbook_1}, \cite{tbook_2}, \cite{nosotros}. In spite of these
arguments, a self gravitating gas of this type of particles accounting for a galactic halo, would have to be
modeled, either as a relativistic MB gas (very light particles  can be relativistic even at low
temperatures) and/or in terms of a distribution that  takes into account Fermi-Dirac or Bose-Einstein
statistics. These studies will be undertaken in future papers \cite{nosotros}.

\begin{acknowledgments}
We thank Professor Rabindra N. Mohapatra for calling our attention to the 
important papers of Ref. \cite{scdm_4} and \cite{scdm_5} and N. Fornengo
for useful discussions. R. A. S. is partly supported by the {\bf DGAPA-UNAM}, 
under grant (Project No. {\tt IN122498}), T. M. is partly supported by 
{\bf CoNaCyT} M\'exico, under grant (Project No. {\tt 34407-E}) and 
L. G. C. R. has been supported in part by the {\bf DGAPA-UNAM} under grant 
(Project No. {\tt IN109001}) and in part by the {\bf CoNaCyT} under grant 
(Project No. {\tt I37307-E}).
\end{acknowledgments}


\end{document}